\documentclass[%
 aip,
 amsmath,amssymb,
 reprint,%
]{revtex4-1}

\usepackage{graphicx}% Include figure files
\usepackage{dcolumn}% Align table columns on decimal point
\usepackage{bm}% bold math
\usepackage[utf8]{inputenc}
\usepackage[T1]{fontenc}
\usepackage{mathptmx}
\usepackage{etoolbox}

\makeatletter
\def\@email#1#2{%
 \endgroup
 \patchcmd{\titleblock@produce}
  {\frontmatter@RRAPformat}
  {\frontmatter@RRAPformat{\produce@RRAP{*#1\href{mailto:#2}{#2}}}\frontmatter@RRAPformat}
  {}{}
}%
\makeatother
\begin{document}

\preprint{AIP/123-QED}

\title{How cytoskeletal crosstalk makes cells move: bridging cell-free and cell studies}

\author{James P. Conboy}
	\thanks{These authors contributed equally to this work.
	}
	\affiliation{
		Department of Bionanoscience, Kavli Institute of Nanoscience Delft, Delft University of Technology, 2629 HZ Delft, The Netherlands
	}%

\author{Irene Ist\'uriz Petitjean}
	\thanks{These authors contributed equally to this work.
	}
	\affiliation{
		Department of Bionanoscience, Kavli Institute of Nanoscience Delft, Delft University of Technology, 2629 HZ Delft, The Netherlands
	}%
 \author{Anouk van der Net}
	\thanks{These authors contributed equally to this work.
	}
	\affiliation{
		Department of Bionanoscience, Kavli Institute of Nanoscience Delft, Delft University of Technology, 2629 HZ Delft, The Netherlands
	}%
 \author{Gijsje H. Koenderink}
 \thanks{These authors contributed equally to this work.
	}
	\email{G.H.Koenderink@tudelft.nl}
	\affiliation{
		Department of Bionanoscience, Kavli Institute of Nanoscience Delft, Delft University of Technology, 2629 HZ Delft, The Netherlands
	}%

\date{\today}% It is always \today, today,

\begin{abstract}
Cell migration is a fundamental process for life and is highly dependent on the dynamical and mechanical properties of the cytoskeleton. Intensive physical and biochemical crosstalk between actin, microtubules, and intermediate filaments ensures their coordination to facilitate and enable migration. In this review we discuss the different mechanical aspects that govern cell migration and provide, for each mechanical aspect, a novel perspective by juxtaposing two complementary approaches to the biophysical study of cytoskeletal crosstalk: live-cell studies (often referred to as top-down studies) and cell-free studies (often referred to as bottom-up studies). We summarize the main findings from both experimental approaches, and we provide our perspective on bridging the two perspectives to address the open questions of how cytoskeletal crosstalk governs cell migration and makes cells move.
\end{abstract}

\maketitle

%%%%

\section{Introduction}

\begin{figure}[h!]
\centering
\includegraphics[scale=0.4]{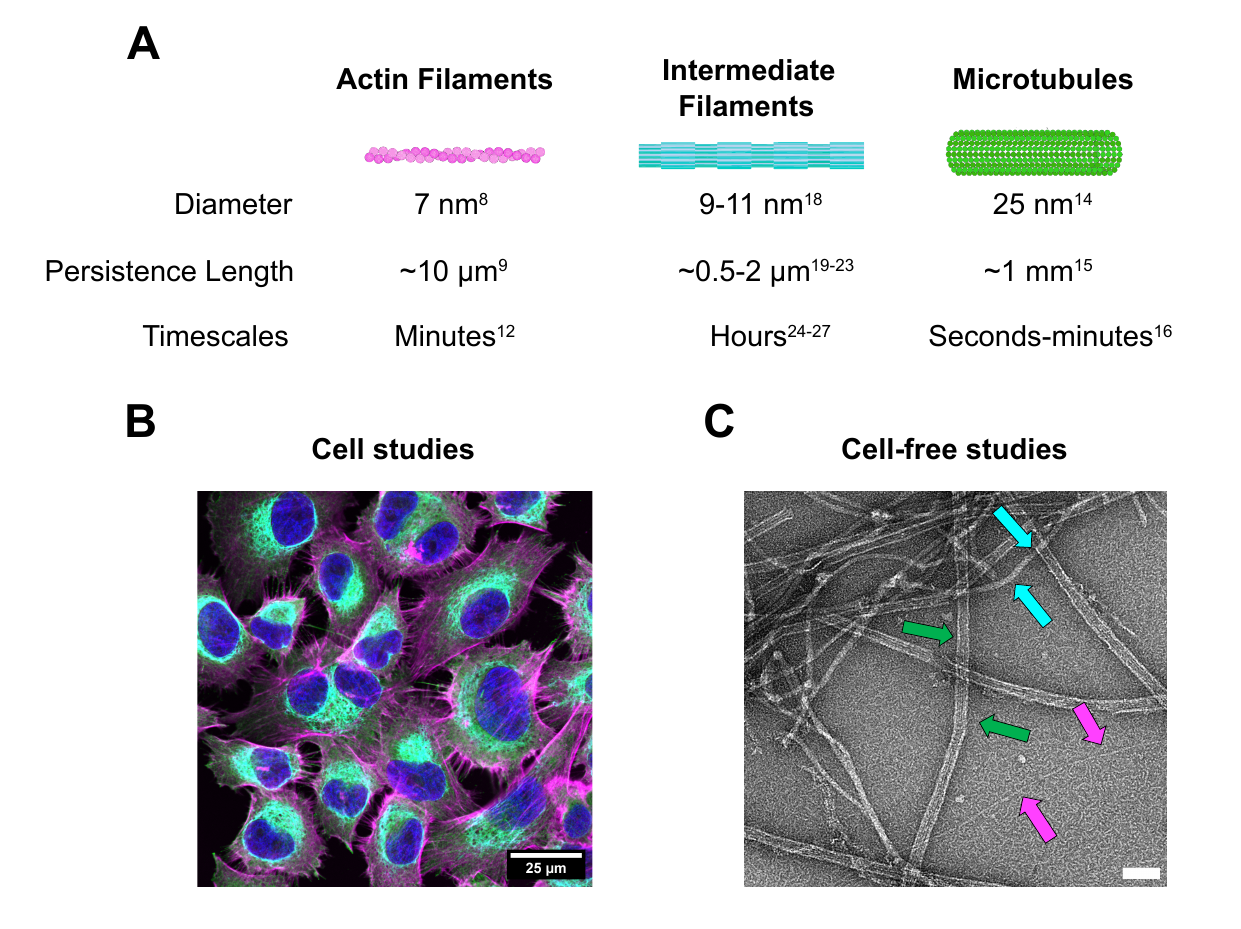}
%\captionsetup{margin=0.5cm}
\caption{(a) Schematic of the three major cytoskeletal filament types and their distinctive physical properties. (b) Fluorescent confocal microscopy image of human melanoma (MV3) cells stained for $\alpha$-tubulin (green), F-actin (red), and vimentin (magenta). The cell nuclei are shown in blue. (c) An electron microscopy image of an \textit{in vitro} reconstituted three-component cytoskeletal network showing F-actin (red arrows), microtubules (green arrows) and vimentin (magenta arrows). Filaments were pre-polymerized separately at 1 $\mu$M. Actin and microtubules were polymerized in MRB0 buffer (80~mM PIPES pH 6.8, 1~mM EGTA and 4~mM MgCl$_2$) with 50 mM KCl, 1 mM DTT and 0.5 mM ATP, while vimentin was polymerized in V-buffer (40~mM PIPES pH 7, 1~mM EGTA and 4~mM MgCl$_2$, 100 mM KCl, 1 mM DTT). The filaments were combined in MRB80 buffer (with 50 mM KCl, 1 mM DTT and 0.5 mM ATP). Scale bar 100 nm.}
\label{f1}
\end{figure}

Cell migration is a process that is fundamental for life. It is a major contributor to tissue morphogenesis in developing embryos \cite{scarpa2016} and drives angiogenesis \cite{lamalice2007}, bone formation \cite{su2018}, tissue repair \cite{fu2019} and immune surveillance \cite{Delgado2022}. On the flip side, however, cell migration is also responsible for pathological cell migration during chronic inflammation \cite{Liu2021-tl} and cancer metastasis \cite{Mierke2019-yw}. Cell migration depends on the mechanical and dynamical properties of the \textit{cytoskeleton}, a network of dynamic biopolymers that self-assemble from small protein building blocks. There are three main cytoskeletal biopolymers: actin filaments, microtubules, and intermediate filaments (Figure \ref{f1}A). They have markedly different structural, mechanical and dynamical properties.

\textit{Actin filaments} are double helices with a diameter of $\sim 7$ nm, made of two strands of globular monomers \cite{Galkin2015-rl}. The filaments are semiflexible since their thermal persistence length $l_p = \kappa / k_BT$ (where $\kappa$ is the bending rigidity and $k_BT$ thermal energy) is $\sim 10 \mu$m, of the same order as the filament contour length \cite{Kang2012-lx}. Actin filaments have an intrinsic structural polarity with a “barbed end” and a “pointed end”. Polymerization-linked ATP hydrolysis causes treadmilling, where the filaments grow at the barbed end and disassemble from the pointed end \cite{Carman2023-fb}. Filaments reconstituted from purified actin turn over slowly (one subunit every 3–4 s), but actin turnover in the cell is catalyzed by actin-binding proteins. Typical actin network turnover times are of order seconds in the leading edge of motile cells \cite{Watanabe2002-nq} to minutes in the actin cortex \cite{Fritzsche2013}. Together with myosin motor proteins, actin filaments form networks and bundles that generate contractile forces \cite{Murrell2015-ur}. \textit{Microtubules} form hollow tubes of 13 protofilaments that are much wider ($\sim$ 25 nm)\cite{Zhang2018-bs} and hence substantially stiffer ($l_p$ $\sim$ mm)\cite{Taute2008-qo} than actin filaments. Like actin filaments, microtubules have an intrinsic structural polarity with distinct plus and minus ends. GTP hydrolysis results in dynamic instability, characterized by alternating phases of microtubule growth and shrinkage \cite{Brouhard2015-uk}. In the cell, this process is tightly regulated by accessory proteins that bind at the microtubule tip or lattice. \textit{Intermediate filaments} are homo-/heteropolymers made of rod-shaped proteins that are encoded by more than 70 genes in humans \cite{Rolleke2023-jb}. Intermediate filament proteins are expressed in a cell-type-specific manner. Mesenchymal cells for instance express vimentin, whereas epithelial cells express keratins. The IF proteins share a common secondary structure consisting of an alpha-helical rod domain flanked by intrinsically disordered head and tail domains. Intermediate filaments are somewhat thicker ($\sim$ 10 nm) \cite{Kang2014} than actin filaments, but they are nevertheless much more flexible ($l_p$ $\sim$ 0.5 - 2 $\mu$m, depending on intermediate filament composition and ionic strength \cite{Beck2010-ip, Noding2012-ee, Pawelzyk2014-qs, Schopferer2009-sk, Mucke2004-ol}) because of their hierarchical rope-like structure. Intermediate filaments are much more stable than actin filaments and microtubules, with slow subunit exchange along their length and annealing and fragmentation on hour time scales in reconstituted systems \cite{noding2014, Tran2023} and in cells \cite{Robert2015-ot, Hookway2015-jf}. 

The physical properties of the cytoskeletal filaments are directly connected to their functions in cell migration. Actin, with its ability to generate protrusive and contractile forces, provides the main driving forces for polymerization-driven \textit{mesenchymal migration} and bleb-based \textit{amoeboid migration} \cite{Rottner2019-ue, Ullo2021-oj}. Meanwhile microtubules play a key role in establishing front-rear polarity and promoting persistent migration, aided by their large persistence length that is much longer than the size of the cell \cite{Akhmanova2022-zz}. Finally, intermediate filaments, with their mechanical resilience, protect the migrating cell and its nucleus from mechanical damage, which is especially important when cells squeeze through confined environments \cite{Patteson2019, Lavenus2020}. 

There is growing evidence that cell migration requires a dynamic interplay between the three cytoskeletal filament systems that depends on mechanical and signaling crosstalk. In mesenchymal migration, coupling of actin to microtubules and intermediate filaments is for instance essential to polarize the actin cytoskeleton and control force generation \cite{Seetharaman2020-ip}. Here we review recent insights in the role of cytoskeletal crosstalk in cell migration, with a focus on mechanical aspects. For more detailed cell biological insights, we refer the reader to several excellent reviews \cite{Dogterom2019, Seetharaman2020, Chung2013, Pimm2021-ti, Schmidt2023}. We take a mainly experimental perspective and refer the reader to other reviews for more theoretically oriented perspectives \cite{Buttenschon2020-bv, Banerjee2019-jc}. Throughout this review, we confront two opposite experimental approaches to studying the biophysics of cytoskeletal crosstalk: \textit{live cell (top-down)} studies (Figure \ref{f1}B) versus \textit{cell-free (bottom-up)} studies of simplified model systems reconstituted from component parts (Figure \ref{f1}C). Live-cell studies have the benefit of physiological relevance, but mechanistic dissection is challenging because of the cell's compositional complexity. Each cytoskeletal system exhibits enormous compositional diversity with different isoforms and posttranslational modifications \cite{MacTaggart2021-pw}. Moreover, cytoskeletal coupling is mechanosensitive as a consequence of mechanosensory signalling loops and transcriptional regulation \cite{Seetharaman2022}. Cell-free studies provide a powerful approach to complement live-cell studies because they allow for highly controlled experiments from the level of single protein, to filaments, to networks. 

\begin{figure}[h!]
\centering
\includegraphics[scale=0.30]{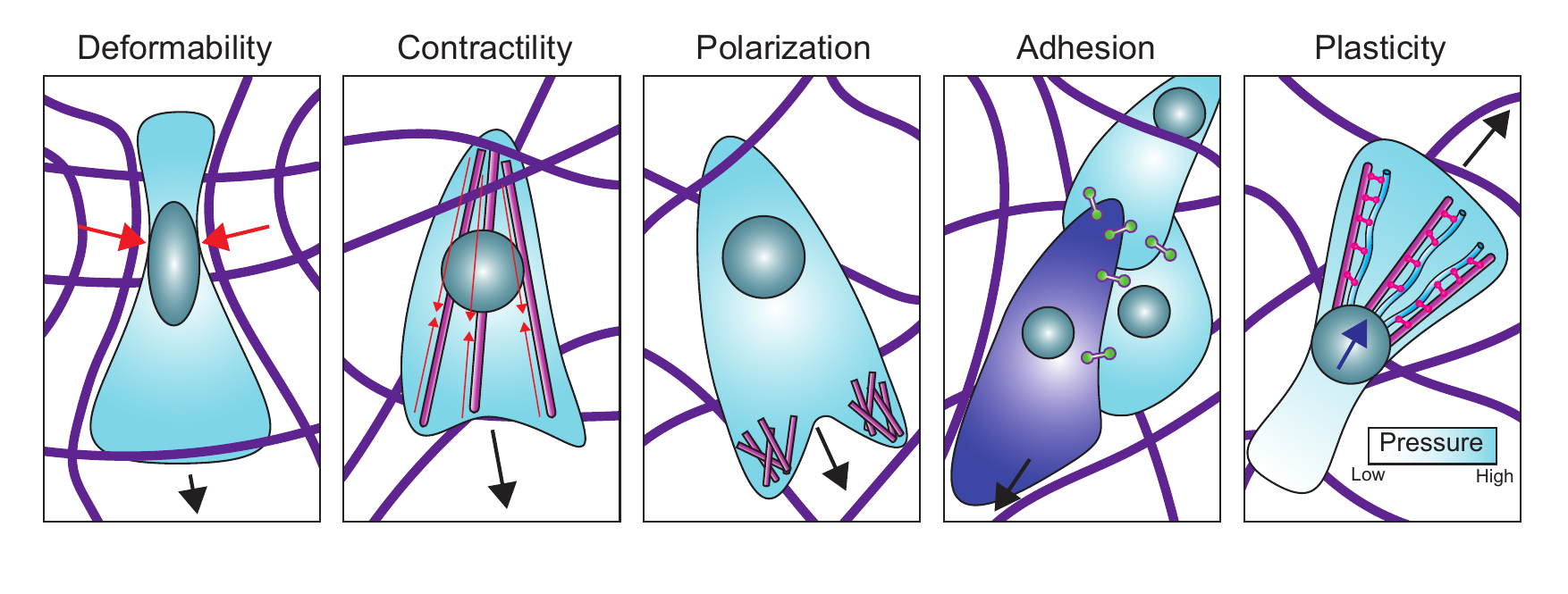}
%\captionsetup{margin=0.5cm}
\caption{Cytoskeletal crosstalk contributes to every aspect of cell migration including: (1) Cell deformability that governs the ability of cells to migrate through confining environments. Red arrows show deformation caused by the cell migrating through the extracellular matrix. (2) Contractility as a major driver of cell motility.  Red arrows show actin-myosin contraction. (3) Front-rear polarity for directional migration. (4) Cell-cell adhesions to coordinate collective migration.  The leader cell is shown in dark blue. (5) Plasticity, the ability of cells to interconvert between different migration strategies in response to their environment, for example between mesenchymal and nuclear piston modes. Here the nucleus is pulled forwards (dark blue arrow).  Black arrows show the direction of migration.  Actin (magenta), vimentin (cyan), nucleus (teal), intercellular adhesions (green linkers), plectin (pink linkers)  and extracellular matrix fibers (purple).}
\label{fig:sample}
\end{figure}

Cytoskeletal crosstalk contributes to every aspect of cell migration (Figure \ref{fig:sample}). We structure the review according to these aspects, from cell deformability, to front-back polarity, contractility, adhesion control in collective cell migration, and finally plasticity, which refers to the ability of cells to adapt their mode of migration to their environment \cite{Te_Boekhorst2016-ax}. We end with a perspective on how connections can be made between cell-free and live-cell studies to address the many open questions on the role of cytoskeletal crosstalk in cell migration.

\section{Cell deformability}

\subsection{Live-cell studies}
Migrating cells must deform their nucleus and cytoskeleton, especially when they move through interstitial tissues that impose significant confinement. Depending on tissue type, cells encounter extracellular matrix (ECM) and interstices between tissues with sizes ranging between 2 and 30 $\mu$m, comparable to their own body  and sometimes even nuclear size \cite{Wolf2009-js}. Metastasizing cancer cells have to overcome even more severe physical barriers as they intravasate across the endothelium into blood vessels or across epithelial tissues into lymphatic vessels. 
Cell deformability is therefore an important determinant of cell migration \cite{Guck2010-xo}. For many cancer cells, for instance, lower stiffness correlates with higher motility \cite{Liu2020-hm}.

Migrating cells experience a complex combination of tensile, compressive and shear deformations of varying amplitude and rate. The effect of these mechanical parameters on cell deformability has been extensively characterized using quantitative biophysical techniques. To study the viscoelastic properties of cells without the impact of cell adhesion, cells can be detached from their substrate and measured in suspension by micropipette aspiration \cite{Gonzalez-Bermudez2019-hu}, optical stretching \cite{Guck2001-qu}, parallel-plates rheometers \cite{Thoumine1997-wn}, or high-throughput microfluidic methods \cite{urbanska2020}. These measurements are mostly relevant for amoeboid migration where cells exhibit only weak adhesion to their environment. For mesenchymal migration, it is more relevant to study mechanics on adherent cells. Whole-cell measurements of adherent cells can be done by monolayer rheology or stretching \cite{Fernandez2007, Sutton2022-es, Bashir2022-en} or by single-cell atomic force microscopy \cite{Beicker2018-cv, Pogoda2022-rc}. Localized measurements to resolve the mechanics of specific subcellular regions can be done by magnetic twisting cytometry \cite{Fabry2001-xm}, atomic force microscopy \cite{Kerdegari2023-ao}, optical or magnetic tweezers \cite{Catala-Castro2022-ha, Bausch1998-ga}, or particle tracking microrheology \cite{Lau2003-lo}. Some of these methods allow for \textit{in situ} measurements in migrating cells \cite{Panorchan2006-bz, Staunton2016-km}.  

The nucleus is the stiffest and largest organelle with a stiffness ranging from 0.1 to 10 kPa, dependent on cell-type \cite{Swift2013-xz}. The main contributors to the rigidity of the nucleus are heterochromatin and the nuclear lamina (also termed the
nucleoskeleton), which contains lamin intermediate filaments \cite{Lammerding2006, Stephens2017-lh, Bell2022-yv}. Intact nuclei \textit{in situ} have a higher stiffness than isolated nuclei because the nucleus is physically coupled to the cytoskeleton through the LINC complex \cite{Liu2014-ub}. Since the nucleus is not only the stiffest but also the largest cellular organelle, it poses a major bottleneck for confined migration \cite{McGregor2016}. When cells are embedded in collagen networks or microfabricated microchannels, their migration velocity linearly decreases with decreasing pore size until migration is physically blocked when the pore size reaches 10\% of the nuclear cross section \cite{wolf2013}. Under highly confined conditions, cells can only move if they are able to remove blocking ECM fibers with proteolytic enzymes \cite{wolf2013, Davidson2014-zh}. 

 The actin cytoskeleton is often considered the main determinant of cell mechanics. Drug-induced depolymerization of actin filaments indeed significantly softens cells both under non-adherent and adherent conditions \cite{Lautenschlager2009-hj,Rotsch2000-tz}. An important contribution of the actin cytoskeleton to cell stiffness comes from contractile forces generated by actin-myosin stress fibers \cite{Martens2008-bc} and by the actin cortex \cite{Fischer-Friedrich2016-oo}. Intermediate filaments form dense networks that are mainly perinuclear, so they contribute little to cortical stiffness but strongly affect the cytoplasmic shear modulus \cite{Guo2013-ue} and the resistance of cells to compression \cite{Pogoda2022-rc}. For leukocytes and tumor cells performing 3D migration, the intermediate filament cytoskeleton is a major determinant of cell deformability. Intermediate filament protein deletion or network disruption causes significant cell softening while at the same time enhancing cell migration \cite{Brown2001-la, Seltmann2013, Patteson2019, Lavenus2020-lj}. Microtubules generally do not contribute much to cell stiffness, with drugs that interfere with microtubule polymerization having minor effects on cell mechanics \cite{Wang1998-yx}. Recently, though, microtubules were shown to exhibit interesting mechano-responsive properties. Cytoskeletal compression induced by cyclic cell stretching or by confined migration was shown to stabilize deformed microtubules by triggering recruitment of CLASP2 \cite{Li2023-kp}. When cells are transferred from rigid 2D substrates to softer 3D hydrogels, the mechanical contribution of microtubules becomes more important because actin stress fibers become less prominent. In cells migrating through collagen gels, microtubules for instance play a crucial role in mechanical support of cellular protrusions \cite{Bouchet2016-dm}. 

It remains an open question how the interactions between the three cytoskeletal biopolymers influence the mechanics of the composite cytoskeleton. Theoretical models predict that composite networks composed of interpenetrating networks of a rigid and a flexible polymer are substantially stiffer than expected from the sum of the moduli of the separate networks \cite{Van_Doorn2017-cp}. Rigid fiber networks by themselves are expected to be soft at low deformation because they deform in a non-affine manner, where the elasticity is governed by fiber bending \cite{Chen2023-va}. The presence of a background network of flexible polymer suppresses these non-affine bending deformations \cite{Van_Doorn2017-cp}. Unfortunately this prediction is difficult to directly test in live-cell experiments because it is very challenging to specifically remove one cytoskeletal network without also affecting the others. Microtubule depolymerization is for instance well-known to activate acto-myosin contraction by the release of the microtubule-associated guanine nucleotide exchange factor GEF-H1 \cite{Chang2008-ix}. At large strains, there is some evidence of mechanical synergy between the cytoskeletal networks. Epithelial cell layers are able to undergo extreme stretching under constant tension ('active superelasticity') by strain-softening of the actin cortex followed by re-stiffening thanks to the keratin intermediate filament network \cite{Latorre2018-bg}. Physical crosslinking between actin and keratin is essential for the maintenance of epithelial stability \cite{Prechova2022}. The ability of flexible polymers to suppress bending deformations of rigid polymers has the interesting consequence that the rigid polymers are reinforced against compressive loads \cite{Das2010-fp}. Under compressive loading, rigid polymers exhibit an Euler buckling instability at a critical compression force $f_c \sim 10\kappa / L^2$, where $L$ is the polymer length. For microtubules, the critical compression force is only of order 1 pN \cite{Dogterom1997-mr}. In the cell, however, microtubules can bear 100-fold larger compression forces because the surrounding actin and intermediate filament cytoskeleton constrains microtubule buckling \cite{Brangwynne2006-nc, Robison2016-ty}. This is consistent with the so-called tensegrity model, which states that cellular shape stability is achieved via a balance between actin filaments and intermediate filaments loaded under tension, and microtubules and thick actin bundles under compression \cite{Wang2001-yc}.

\subsection{Cell-free studies}

Live-cell mechanical measurements can be difficult to interpret in quantitative terms because they are sensitive to the amplitude, type and rate of deformation, geometry of the mechanical probe, the probed location in the cell, and the cell's extracellular environment \cite{Wu2018}. Cell-free studies provide a useful complement because they permit quantitative measurements of the mechanical properties of isolated cytoskeletal components, both at the single filament and at the network level. 

At the single filament level, cytoskeletal biopolymers have been bent, stretched, compressed, and twisted using optical and magnetic tweezers \cite{Block2018, Van_Mameren2009-mb, Yasuda1996-hr, Bibeau2023-nm}, atomic force microscopy \cite{Kreplak2005, Kis2002-xb}, and microfluidic flow devices \cite{Schaedel2015-uc}. Actin filaments and microtubules have a high bending and stretching rigidity, but they break at rather low tensile strains ($\sim$ 150\% strains) \cite{Tsuda1996-ru, Endow2019-we}. Moreover, actin filaments become more fragile under torsion \cite{Tsuda1996-ru} and microtubules soften upon repeated bending \cite{Schaedel2015-uc}. This fragility is likely related to the fact that actin filaments and microtubules are made of globular subunits. By contrast, intermediate filaments are made of fibrous subunits held together by extensive lateral interactions. Intermediate filaments easily stretch and bend due to their open structure and they can withstand tensile strains of more than 200$\%$ before rupture \cite{Kreplak2005}. Similar to a car's safety belt, intermediate filaments are soft under small and slow deformations but stiff under large and fast deformations \cite{Block2018}. Recent evidence suggests that different intermediate filament proteins respond differently to tensile loads. When subjected to stretch-relax cycles, keratin filaments elongate with every cycle but keep the same stiffness, whereas vimentin filaments soften with every cycle but always return to the same initial length \cite{Lorenz2023-ll}. It appears that vimentin stretches by monomer unfolding \cite{Nunes_Vicente2022-fh}, whereas keratin filaments stretch by viscous sliding of subunits \cite{Lorenz2023-ll}. It will be interesting to see what further diversity may be generated by co-polymerization of different intermediate filament proteins and by post-translational modifications. 

On the network level, mechanical properties of cytoskeletal filaments are most conveniently probed by either bulk rheology or microrheology. In bulk rheology, cytoskeletal networks are sheared between the two parallel plates of a rheometer, providing a read-out of the macroscopic viscoelastic response \cite{Gardel2008-uo}. Microrheology instead probes the localized viscoelastic response of a material by tracking the motion of embedded probe particles, either in response to thermal fluctuations (passive microrheology \cite{Weihs2006-uv}) or to a force applied by optical or magnetic tweezers (active microrheology \cite{Yang2013-pv}). The mechanical response of cytoskeletal networks is determined by an interplay of the stiffness of the filaments and their interactions. 

Actin filaments and microtubules form entangled networks that easily fluidize under shear due to filament disentanglement \cite{Lin2007, Yang2012, Kirchenbuechler2014-dh}. Filament crosslinking prevents this fluidization and causes the networks to strain-stiffen. This strain-stiffening response is only moderate for microtubules because of their high rigidity and because shearing causes force-induced unbinding of crosslinks \cite{Lin2007, Yang2013-pv}. Actin networks exhibit more pronounced strain-stiffening because their elasticity is affected by the entropic elastic response of the filaments to tensile loading \cite{Gardel2004}. Tensile loading reduces the conformational entropy of actin filaments, pulling out bending fluctuations, causing entropic strain-stiffening \cite{MacKintosh1995-fk}. Increased crosslink densities shift the onset of strain-stiffening to smaller shear strains because less excess length is stored in bending fluctuations when the crosslinks are more closely spaced \cite{Gardel2004}. Some crosslinker proteins (most notably filamin) are so large that their compliance directly contributes to the network response. Crosslinker extensibility increases the rupture strain by postponing the point where the actin filaments experience tensile loading \cite{kasza2010, broedersz2010}. Bundling of actin filaments, which is common at high concentrations of crosslinker proteins, suppresses entropic elasticity. Bundled actin networks still strain-stiffen \cite{Gardel2004}, but by an enthalpic mechanism that involves a transition from soft bending modes at low strains to rigid stretching modes at high strain \cite{Wilhelm2003-ry, Head2003-pm}. Under compression, actin and microtubule networks soften due to filament buckling \cite{Pogoda2022-rc}. For branched actin networks, compressive softening has been shown to be reversible, likely because the buckled filaments are prevented from collapsing by their connections with the network \cite{Chaudhuri2007-uf}.

The mechanical properties of intermediate filament networks differ in various respects from those of actin and microtubule networks. First, intermediate filaments form strain-stiffening networks even in absence of any crosslinker proteins, as demonstrated for vimentin, neurofilaments, desmin, and keratin \cite{Lin2010, yao2010}. The filaments spontaneously form crosslinks mediated by electrostatic interactions between their disordered C-terminal tails. Upon tail truncation, the networks no longer strain-stiffen \cite{Lin2010, Aufderhorst-Roberts2019, Bar2010-ji}. The effective crosslink density depends on the concentration of divalent cations such $Mg^{2+}, Ca^{2+}$ or $Zn^{2+}$ \cite{Lin2010, yao2010, Wu2020, Fukuyama1978} and is sensitive to the buffer ionic strength and pH \cite{Schepers2021}. For keratins, there are additional hydrophobic interactions between the central rod domains that stiffen the networks \cite{Pawelzyk2014-qs}. Second, intermediate filament networks have much larger rupture strains than actin and microtubule networks as a consequence of the larger single-filament extensibility. This is reflected in the dependence of the elastic modulus $K$ on the applied shear stress $\sigma$. While actin networks only exhibit an entropic strain-stiffening regime where $K$ increases as $\sigma^\frac{3}{2}$, intermediate filament networks exhibit an additional enthalpic regime where $K$ increases more weakly, reflecting strain-induced filament alignment \cite{Lin2010}. After yielding, intermediate filament networks can even recover their initial shear modulus, likely by the re-establishment of tail-tail crosslinks \cite{Wagner2007, Aufderhorst-Roberts2019}.

Recently there has been increasing attention for the mechanical properties of cytoskeletal composites. Reconstitution of composite networks requires careful tuning of the buffer conditions since the different cytoskeletal polymers are traditionally reconstituted in their own optimized buffer conditions. Intermediate filaments are especially sensitive to solution pH and ionic concentrations because they are prone to polymorphism \cite{Denz2021-kk}. Until now nearly all studies of composite networks have focused on two-component composites of cytoskeletal filaments co-polymerized in the absence of crosslinkers. At small strains, co-entangled composites (specifically combinations of actin/vimentin \cite{Golde2018}, actin/keratin \cite{Deek2018-hv, Elbalasy2021-zb}, actin/microtubules \cite{Pelletier2009-en}, and vimentin/microtubules \cite{34152a02b0334a69ac3014e404b46e11} have generally been shown to exhibit a simple additive viscoelastic response. However, there is evidence for direct interactions of vimentin filaments with actin filaments \cite{Esue2006} as well as microtubules \cite{Schaedel2021}, which could potentially influence the network rheology. These interactions could potentially lead to cell-type specific cytoskeletal crosstalk, since they are mediated by the C-terminal tail of intermediate filaments that shows large length and sequence variations between different intermediate filament proteins. It was furthermore shown that vimentin can impose steric constraints that hamper actin network formation and thus cause network weakening \cite{Jensen2014-cs}. At large strains, there is evidence of synergistic enhancement of the mechanical properties in certain cytoskeletal composites. For actin/keratin composites, the strong strain-stiffening response of the keratin network was found to dominate the high-strain response of the composites \cite{Elbalasy2021-zb}. For actin/microtubule composites, microtubules were shown to promote strain-stiffening of the actin networks, even at low density \cite{Lin2011, Ricketts2018-ie}. This effect was explained by the ability of rigid microtubules to suppress nonaffine bending fluctuations of actin filaments. It will be interesting to explore how these synergies are modified in the presence of crosslinkers. Recent work showed that when actin filaments and microtubules are crosslinked to each other by biotin-streptavidin, the composite is more elastic than when both filaments are independently crosslinked \cite{Ricketts2019-it}. 

To the best of our knowledge, there has so far been only one study of three-component networks combining actin, vimentin and microtubules \cite{Shen2021}. It was shown by microrheology that the linear elastic modulus of the composite is dominated by actin, with little contribution from either microtubules or vimentin. Yet vimentin was shown to significantly extend the elastic regime to longer timescales. The authors proposed that the vimentin network that fills in the pore spaces of the actin network \cite{wu2022} slows stress relaxation by constraining actin reptation. More work is needed to systematically study cytoskeletal composites and to explore the impact of crosslinking with cytolinker proteins such as plectin. Due to their high molecular weight these proteins are difficult to purify. To circumvent this problem, one can engineer proteins that contain only the cytoskeletal binding proteins separated by a spacer \cite{Preciado_Lopez2014-af}. Using this approach, we recently found that crosslinking with a plectin-mimetic crosslinker causes synergistic stiffening of actin-vimentin composites \cite{Petitjean2023}. 

\section{Cell contractility}

\subsection{Live-cell studies}

The actin cytoskeleton is the engine behind cell migration\cite{SenGupta2021-ht}.~Depending on the extracellular environment, cells can switch between different mechanisms that use actin-based forces in different ways\cite{Petrie2016}. Fibroblasts and other adherent cells perform \textit{mesenchymal migration}, which relies on integrin-based adhesion to the extracellular matrix (ECM). The process occurs via a four-step cycle. First, actin polymerization pushes against the membrane at the leading edge, producing lamellipodia in cells migrating on flat rigid surfaces or pseudopodia in cells migrating in 3D extracellular matrices. Next, the cell generates integrin-based adhesions with the substrate that connect to the contractile machinery of acto-myosin stress fibres. Through a combination of pulling from the front and squeezing from the rear, the cell body moves forward. Finally, old adhesions are detached from the substrate or dissolved at the trailing edge \cite{Palecek1998}. The contractile forces involved in cell migration have been measured through the traction forces exerted on the substrate. This is usually done by adhering cells to a hydrogel substrate with known mechanical properties, such as polyacrylamide. By measuring the displacements of fluorescent tracer particles incorporated in the gel with fluorescence microscopy, one can computationally infer the traction forces using continuum mechanics models \cite{Style2014}. Adherent cells that experience strong confinement utilize a \textit{nuclear piston} mechanism where actin-myosin contraction in front of the nucleus pulls the nucleus forward. Since the nucleus divides the cell in forward and rearward compartments, it acts as a piston that pressurizes the forward compartment and drives forward a cylindrical lobopodial protrusion \cite{Petrie2014-zo}. Weakly adherent cells such as leukocytes and physically confined fibroblasts and cancer cells perform \textit{amoeboid migration}, characterized by spherical membrane blebs at the leading edge (reviewed in \cite{Paluch2013-qs}). Blebs are created by myosin-driven contraction of the actin cortex underneath the cell membrane, which builds up hydrostatic pressure in the cytoplasm. Local rupture of the actin cortex or its attachment to the membrane causes local membrane delamination, pushing forward a membrane bleb. Over time the actin cortex regrows under the bleb membrane and myosin contraction drives bleb retraction. Confinement can also induce other migration modes that require little substrate adhesion. Cells can move via friction generated by actin flows within the cortex generated by myosin contraction and actin turnover \cite{Liu2015-hp}, and some tumor cells can still migrate by using active transport of water from the front to the back of the cell to propel themselves forward (\textit{osmotic engine} model \cite{Stroka2014-xy}).

While not being components of the contractile machinery, both microtubules and intermediate filaments are important for regulating cell contraction. Microtubules negatively regulate the assembly and contractility of actin stress fibers by sequestering GEF-H1, an activator of the small GTPase Rho, in an autoinhibited state \cite{Krendel2002-yw}. Microtubule depolymerization by nocodazole releases active GEF-H1, leading to a global increase of contractility as measured by traction force microscopy \cite{Rape2011}. During both mesenchymal and amoeboid migration, microtubule depolymerization and consequent GEF-H1 is tightly regulated so that actin contractility can be precisely timed and localized in a mechanosensitive manner \cite{Azoitei2019-ge, Heck2012-aa, Kopf2020-eu}. Besides biochemical regulation, it is likely that mechanical synergy is also involved in microtubule-based control of actin contractility, since microtubules are able to absorb some of the forces from the contractile actin cytoskeleton \cite{Wang2001-yc}. 

Intermediate filaments likewise regulate actin-based cell contraction by a combination of mechanical synergy and biochemical signaling. In cells migrating on flat surfaces, vimentin has been reported to inhibit stress fiber assembly and contractility through down-regulating GEF-H1 and RhoA \cite{Jiu2017}. Nevertheless, traction force measurements have shown that vimentin-null cells are less contractile than their wild-type counterparts \cite{wu2022}. Taken together with the observation that vimentin filaments orient traction stresses along the front-rear axis, this suggests a mechanical synergy where vimentin helps build up and transmit larger contractile forces \cite{Burckhardt2016}. Recently it was shown by structured illumination microscopy and electron microscopy that vimentin filaments are closely associated with actin stress fibers, forming meshworks that wrap around stress fibers or co-align with them \cite{wu2022, Seetharaman2022}. Physical coupling between the two systems is dependent on the cytolinker protein plectin \cite{Jiu2015} Interestingly, it was recently shown that plectin binds vimentin in response to acto-myosin pulling forces \cite{Marks2022}. The mechanism for this mechanosensitivity is unknown but could involve catch bonding\cite{Daday2017}. Plectin-mediated coupling of actin and vimentin was recently shown to be essential for cells migrating via the nuclear piston mechanism \cite{Marks2022}. The vimentin network helps transmit acto-myosin pulling forces to the nucleus, thus enhancing the pressure in the front of the nucleus. It is not yet known whether intermediate filaments also influence cell migration modes driven by contractile activity of the actin cortex, but recent observations that vimentin and F-actin are associated within the cell cortex suggest this is likely \cite{wu2022}.

\subsection{Cell-free reconstitution studies}
There is an extensive body of work using cell-free reconstitution to elucidate the mechanisms by which myosin II motor proteins contract actin networks (reviewed in \cite{Murrell2015-ur}). The contraction mechanism has been found to depend on the actin network connectivity, which is controlled by filament length and by crosslinking. Well-connected networks of long filaments contract because myosins generate compressive stress that causes the actin filaments to buckle and break \cite{Murrell2012-sh}. By contrast, when the filaments are short, myosins contract the network by polarity sorting, transporting and clustering actin filament plus ends to form polar actin asters \cite{Wollrab2018-zo}. In both cases, the length scale of contraction is set by the network connectivity. Global network contraction requires the actin network to be crosslinked above a critical percolation threshold \cite{Alvarado2017}. However, excessive crosslinking will prevent contraction by making the network too rigid \cite{Bendix2008-px}. As described above, several cell migration mechanisms rely on myosin-driven contraction of the actin cortex. Recently several groups have been able to reconstitute biomimetic actin cortices by co-encapsulating actin and myosin inside cell-sized lipid vesicles. For weak actin-membrane attachment, the network detaches from the membrane upon contraction \cite{Carvalho2013}. In case of stronger attachment, myosin contraction can cause membrane blebbing \cite{Loiseau2016-tu}. Cortical flows that are important for driving amoeboid migration require not only myosin activity, but also network remodeling through actin depolymerization \cite{McFadden2017-fq}. Under particular conditions, crosslinked actin-myosin cortical networks in emulsion droplets have been observed to exhibit cortical flows  \cite{Vogel2020-dh}, likely because myosin can promote actin turnover \cite{sonal2019}. Cell extracts, which contain additional proteins to promote actin turnover, also exhibit cortical flows when encapsulated in emulsion droplets \cite{Tan2018, Pinot2012-dw, Malik-Garbi2019-xd}. When these droplets are confined, the myosin-driven cortical flows can propel the droplets forward due to friction with the channel walls, mimicking amoeboid migration of nonadhesive cells \cite{Sakamoto2022-hi}.

So far only few studies have looked at the effect of intermediate filaments or microtubules on contraction of actin-myosin networks. The addition of a vimentin network that interpenetrates an actin network has been shown to promote myosin-driven contraction by increasing the network connectivity \cite{Shen2021a}. Similarly, also the addition of microtubules has been shown to promote uniform macroscopic myosin-driven contraction \cite{Lee2021}.  

\section{Front-rear polarization}

\subsection{Live-cell studies}
Directed cell migration requires the breaking of cell symmetry to generate a cell front and a cell rear along an axis aligned with the direction of locomotion. Until now, the role of cytoskeletal crosstalk in front-rear polarity has mostly been studied in the context of 2D mesenchymal cell migration \cite{Cramer2010-fi}. It is long known that the microtubule cytoskeleton is essential for maintaining a polarized distribution of actin-based forces with actin polymerization in the front and myosin II-based contraction forces in the cell body and rear \cite{Vasiliev1970}. Microtubules align along the axis of cell movement with their plus ends oriented towards the leading edge. They appear to stimulate actin-driven cell protrusion by multiple mechanisms. They activate Rac1 and inhibit Rho, therefore promoting actin polymerization and preventing myosin-II-driven contractility at the leading edge. Moreover, actin filaments have been observed to grow directly from microtubule tips toward the leading edge in growth cones of neurons, with the help of protein complexes involving APC and CLIP-170 \cite{Efimova2020-nb, Henty-Ridilla2016-zp}. There is an interesting actin/microtubule reciprocity, though, since the microtubules require guidance along actin stress fibers to reach the leading edge. 
This guidance requires actin-microtubule crosslinking, for instance by ACF7, Growth Arrest-Specific Proteins (Gas2L1), CLIP-associating proteins (CLASPs) or drebrins (reviewed in \cite{Dogterom2019}). These proteins target growing microtubule plus ends by binding to EB (end-binding) proteins, and all of them except drebrin also possess a microtubule-lattice-binding domain. When these crosslinkers are depleted from cells, microtubules cease to grow along actin stress fibers and the microtubule array loses its front-rear polarity \cite{Kodama2003, Drabek2006}. Persistent cell migration is strongly hampered as a consequence, not only because actin-based protrusions is misregulated, but also because microtubules fail to reach cortical microtubule stabilizing complexes (CMSCs) that surround focal adhesions \cite{Bouchet2016-fh}. Tethering and stabilization of microtubule plus ends by CMSC binding is required for microtubule-dependent focal adhesion turnover, which is essential for migration (reviewed in \cite{Schmidt2023}). It is not yet clear how these crosstalk mechanisms are modified when cells perform 3D mesenchymal migration, but likely the core mechanisms are shared. One important new factor in 3D migration is that microtubules have a more important mechanical role and are needed to support pseudopodia \cite{Bouchet2016-dm}. A second important new factor is that the rigidity of the nucleus hampers migration through small pores. It was recently shown that microtubules anchored to the nucleus play an important role in active transport of MT1-MMP to the cell surface where it drives extracellular matrix proteolysis in front of the nucleus \cite{Infante2018-ai}. 

Although intermediate filaments lack intrinsic polarity, they do contribute to directed mesenchymal migration \cite{Chung2013}. When the vimentin network is disassembled using peptides or when vimentin expression is knocked down, cells lose their polarity and lamellipodia appear all around the cell \cite{Helfand2011}. Vimentin forms closely associated parallel arrays with microtubules in migrating cells \cite{Sakamoto2013-jv, Gan2016-jd}. Experiments conducted using vimentin-deficient mouse embryonic fibroblasts attached to polarized and non-polarized protein micropatterns demonstrated that the lack of vimentin alters microtubule organisation, disrupting cell polarity \cite{Goldman2014}. The two cytoskeletal networks organize in an interdependent manner. The vimentin distribution is polarized by a collaboration between active motor-driven transport along microtubules and actin-driven retrograde flow \cite{Leduc2017-rl}. Conversely, since the vimentin network is about 10-fold more long-lived than the microtubule network, it can serve as a template for guiding microtubule growth along previous microtubule tracks \cite{Burckhardt2016}. This provides a feedback mechanism to sustain front-rear polarity. Moreover, the alignment of the vimentin network with the polarity axis mechanically integrates actin-based forces and orients them to promote directional migration \cite{Costigliola2017}. This mechanical integration is probably aided by vimentin-microtubule crosslinker proteins such as plectin and APC \cite{Sakamoto2013-wt}. In addition to this mechanical role, there is growing evidence for signalling functions of intermediate filaments in cell migration (reviewed in\cite{Seetharaman2020}). At the cell periphery, there is for instance Rac-mediated crosstalk between vimentin and actin, where Rac causes vimentin disassembly by phosphorylation of vimentin, promoting actin-driven membrane protrusion \cite{Helfand2011}. Intermediate filaments also regulate focal adhesion clustering and turnover by binding integrins and via biochemical signalling \cite{Leube2015-dp}.

\subsection{Cell-free reconstitution studies}
Several studies have explored how interactions between two different cytoskeletal filament types may contribute to the front-rear polarity of migrating cells. These studies mostly used surface assays where one or both cytoskeletal filaments were surface-anchored to facilitate imaging and control the geometry of interaction. Just a few of these investigated the interplay of intermediate filaments with actin or microtubules. When surface-anchored microtubules are grown in the presence of an entangled vimentin network, they were found to be stabilized against depolymerization by direct interactions with vimentin filaments \cite{Schaedel2021}. Vimentin attachment reduced the catastrophe frequency and induced rescue of depolymerizing microtubules. However, in the absence of crosslinker proteins, these interactions were found to be short and infrequent. It is likely that vimentin-microtubule crosslinkers such as APC and plectin create more drastic effects on vimentin and microtubule polymerization. Interestingly, the vimentin-binding region of APC by itself was shown to promote vimentin polymerization \cite{Sakamoto2013-wt}, which may perhaps promote vimentin polymerization along microtubules. It was recently shown that actin and vimentin filaments do not interact in the absence of crosslinkers, but when an engineered plectin-mimicking crosslinker was added, actin filaments polymerized along surface-anchored vimentin filaments\cite{Petitjean2023}.  

A larger set of studies investigated the interplay of microtubules with different actin network structures designed to mimic structures found at front of crawling cells. Branched or densely entangled actin network that mimic the dense actin array in the lamellipodium were shown to act as a steric barier for microtubule growth\cite{Preciado_Lopez2014-xm, Inoue2019-ky, Gelin2023-fj}. However, when microtubules were crosslinked to actin by Tau protein, they were able to generate sufficient polymerization force to penetrate dense actin barriers \cite{Gelin2023-fj}. By contrast, when actin was arranged in stiff bundles that mimic actin stress fibers and bundles in filopodia, steric interactions were instead found to promote alignment and growth of microtubules along the actin bundles \cite{Preciado_Lopez2014-xm, Gelin2023-fj}. Actin-microtule crosslinking proteins such as ACF7, Gas2L1, or CLASP2 were shown to promote actin-guided microtubule growth by allowing growing microtubules to be captured by and zippered along the actin bundles \cite{Preciado_Lopez2014-xm, Gelin2023-fj, Van_de_Willige2019-pu, Elie2015-mw, Preciado_Lopez2014-af, Alkemade_thesis}. Conversely, microtubules can also influence actin polymerization. Microtubule-lattice binding crosslinkers can induce guided polymerization of actin filaments along microtubule \cite{Elie2015-mw, Rodgers2023-lh}. Microtubule-tip binding crosslinkers can induce active transport of actin filaments by the growing microtubule tip \cite{Preciado_Lopez2014-af, Alkemade2022}. Computer simulations and theoretical modeling showed that this transport is driven by the affinity of the cross-linker for the chemically distinct microtubule tip region\cite{Alkemade2022}. These interactions may potentially enable growing microtubules to relocate newly nucleated actin filaments to the leading edge of the cell and thus boost migration. Altogether, these studies suggest that coupled polarization of the three cytoskeletal filament systems can at least partly be understood on the basis of a mechanical interplay.

\section{Collective Migration \& Intercellular Adhesions}

\subsection{Live-cell studies}

Many cell types have the ability to synchronize their movement and perform collective migration. Collective migration is important for organogenesis and wound healing but also contributes to cancer metastasis. Depending on cell type and tissue context, different modes of collective migration can emerge. Epithelial cells tend to move as sheets adhered to the extracellular matrix \cite{Vishwakarma2020-of}, while cancer cells often migrate as three-dimensional strands or clusters through tissues \cite{Haeger2015}. Remarkably, multi-cellular migrating structures behave similarly to liquid crystalline materials and undergo solid-to-liquid transitions in response to confinement. These jamming/unjamming transitions are linked to cell and nucleus shapes \cite{Grosser2021, Park2015} and are determined by molecular interactions that regulate cell-matrix and cell-cell adhesions \cite{Ilina2020, Mongera2018, Palamidessi2019}. Traction force measurements for epithelial and endothelial monolayers have shown that cells within the monolayer tend to migrate in the direction in which the normal stress is greatest and the shear stress least \cite{Tambe2011}. This mechanism of collective cell guidance called \textit{plithotaxis} critically relies on mechanical coupling between the cells by cell-cell adhesions. Plithotaxis is regulated by the tumor suppressor protein merlin, which coordinates polarized Rac1 activation and lamellipodium formation at the multicellular scale \cite{Das2015}. We speculate that, since Rac1 is an important shared regulator of all three cytoskeletal systems, there could be crosstalk with intermediate filaments and microtubules in plithotaxis. At the same time, intercellular adhesions help collectively migrating cells to establish supracellular polarization with leader cells at the front and follower cells behind \cite{Campas2023}. The leader cells explore the tissue environment using focal adhesions, find the path, and - if necessary - carve out a path by degrading the ECM. Cancer cells dynamically rearrange leader and follower positions during collective invasion to cope with the large energy usage of the leader cells\cite{Zhang2019}.

Epithelial and endothelial cells interact through mechanosensitive \textit{adherens junctions} based on classical cadherins and VE-cadherins, respectively, which connect to the actin cytoskeleton via $\alpha$-catenin and vinculin\cite{Campas2023, Broussard2020-qu}. Endothelial cells are additionally connected by complexus adherens junctions that connect to vimentin via VE-cadherin \cite{Schmelz1993-pe}. Epithelial cells are additionally connected by \textit{desmosomes} based on desmosomal cadherins that connect to the keratin intermediate filament cytoskeleton via the adaptor proteins desmoplakin, plakophilin, and plakoglobin \cite{Hegazy2022-bi, Lewis1997, Broussard2020-qu}. The desmosomal junctions are expendable under homeostatic conditions but strictly required for preserving cell-cell adhesion under mechanical stress \cite{Sadhanasatish2023-tl}. 
There is growing evidence that cytoskeletal crosstalk is important for the organization, dynamics and mechanoresponsiveness of intercellular adhesions during multicellular migration, although the exact extent is not clear. 

Keratin intermediate filaments and actin jointly influence desmosomes and adherens junctions because they are crosslinked via plectin. In epithelia, plectin organizes keratin into a rim-and-spoke configuration where contractile forces generated by acto-myosin are balanced by compressive elements provided by the keratin network, thus balancing internal tension and stabilizing cell-cell contacts \cite{Prechova2022}. Deletion of plectin therefore causes perturbations of both desmosomes and adherens junctions. Although endothelial cells do not have desmosmes, also here plectin-mediated crosslinking between F-actin and vimentin intermediate filaments regulates adherens junction strength and tissue integrity \cite{Osmanagic-Myers2015}. Migrating epithelial cells need to dynamically rearrange their adhesive contacts. Desmosome remodeling is dependent on both actin and keratin. Assembly of desmosomes at the leading edge and subsequent transport to the lateral sides is mediated through extensive actin remodelling, while more matured desmosomes are guided via keratins to the cell center to eventually disassemble \cite{Roberts2011}. Intermediate filaments have an important mechanical role in distributing actin-myosin based forces, similar to their role in single-cell migration. Collective movement of both embryonic cells and astrocytes with proper leader-follower dynamics was shown to be highly dependent on this mechanical synergy of actin and intermediate filaments \cite{Sonavane2017-ng, DePascalis2018}. 3D tumor cultures of $ex$ $vivo$ breast cancer demonstrated heterotypic keratin expression between leader and follower cells\cite{Cheung2013}, which was speculated to be necessary to regulate individual cytoplasmic viscoelasticity and mechanical coupling through desmosome anchoring during collective invasion\cite{Yoon2019}. Invasion assays of epithelial cancers with manipulated keratin expressions also indicate that keratin expression can regulate migration mode; positive keratin-14 cells are leaders of migrating strands in organoid carcinoma assays, while keratin-8 and keratin-18 depletion can shift migration from individual to collective \cite{Cheung2013, Fortier2013}. Interestingly, keratins can also organize asymmetrically in migrating cells and localize in lamellipodia to support polarization and invasive phenotype, which is mediated by actin filaments \cite{Kolsch2009}.

There is ample evidence that microtubule-actin crosstalk affects adherens junctions via mechanisms similar to those observed for focal adhesions. Adherens junctions contain multiple proteins that bind microtubule plus ends, including APC, ACF7 and CLASP \cite{Vasileva2018-lc, Shahbazi2013}. This allows microtubules to promote myosin II activation and local concentration of cadherin molecules \cite{Stehbens2006} and facilitates trafficking of junctional components to the cell surface \cite{Chen2003-ob, Ligon2007}. At the same time, microtubules promote junctional actin assembly by promoting liquid-liquid-phase separation of the actin nucleator cordon bleu (Cobl) \cite{Tsukita2023-vr}. During collective migration in vivo, it was was found that cell-cell contacts differed in their requirement for dynamic microtubules along the leader-follower axis \cite{Revenu2014}. Cells of the leading domain remained cohesive in the absence of dynamic microtubules, whereas dynamic microtubules were essential for the conversion of leader cells to epithelial followers. Interestingly, it was recently shown that physical confinement of collectively migrating cancer cells can induce the dissociation of leader cells by actin-microtubule crosstalk \cite{Law2023-bs}. Confinement-induced microtubule destabilization releases and activates GEF-H1, which promotes RhoA activation and results in leader cell detachment.

\subsection{Cell-free reconstitution studies}

Interactions between cytoskeletal filaments and cell-cell adhesion complexes are relatively unexplored in reconstituted systems. Adherens junctions are based on transmembrane cadherin adhesion receptors. The extracellular domains of cadherins on adjacent cells form adhesions by forming homodimers. The cytosolic domains of cadherins bind $\beta$-catenin, which in turn binds $\alpha$-catenin \cite{Maker2022}. Biochemical studies showed that actin filaments do not bind directly to cadherins, but are tethered indirectly via $\alpha$-catenin \cite{Rimm1995, Rangarajan2023-hy}. In solution, $\alpha$-catenin does not interact with actin filaments and the cadherin-$\beta$-catenin complex simultaneously \cite{Yamada2005}. However, under tension $\alpha$-catenin stably connects the cadherin-catenin complex to actin filaments via a directionally asymmetric catch bond \cite{Buckley2014, Arbore2022-on}. This mechanosensitivity implies directional regulation of cell-cell adhesion in response to tension, which may connect to cooperative phenomena such as (un)jamming and plithotaxis mentioned above. Biochemical studies have shown that $\alpha$-catenin influences the organization of junctional actin both directly, by inhibiting barbed-end growth, and indirectly, by interactions with various actin-binding proteins including Arp2/3 \cite{Hansen2013}. With the complexity of cadherin-actin interactions starting to become uncovered, it will be interesting to study how crosstalk with microtubules may further regulate adherens junctions. Recently a first study in this direction was able to reconstitute the effect of microtubules on junctional actin nucleation, as was described above in live-cell studies  \cite{Tsukita2023-vr}. When dynamic microtubules were incubated together with G-actin and Cobl, actin filaments were found to be nucleated via Cobl condensates from both the tips and lattice of microtubules. 

Regarding desmosomes, so far there have been mainly biochemical and structural studies, but few cell-free biophysical studies. Biochemical studies of the desmosomal cadherins showed that they form cell-cell connections via heterodimerization of the extracellular domains of desmogleins and desmocollins \cite{Harrison2016-tl}. Single-molecule force spectroscopy by AFM showed that the binding force is 30 to 40 pN \cite{Fuchs2022-vl}. The cytoplasmic domains of desmosomal cadherins bind plakoglobin and plakophilins, an interaction that has also been reconstituted \cite{Mathur1994-zl}. Plakoglobin and plakophilins in turn bind to desmoplakin \cite{Kowalczyk1997-wz}, which finally binds keratins. Biochemical and structural studies showed that desmoplakin interacts via its C-terminus to the rod domain of keratin \cite{Fontao2003-xw, Favre2018-sw}. To the best of our knowledge, there are no cell-free studies addressing cytoskeletal crosstalk with desmosomes.

\section{Plasticity of cell migration}

\subsection{Live-cell studies}
Most terminally differentiated cells such as epithelial and stromal cells migrate only during morphogenesis. However, tissue injury can induce cell plasticity. Mature cells can reenter the cell cycle and change their phenotype guided by paligenosis programs \cite{Brown2021}. Unfortunately plasticity can also contribute to disease. For instance, malignant cancer cells are often hyperplastic, contributing to their invasiveness. A well-studied example of cell plasticity is epithelial-mesenchymal transition (EMT), a reversible process in which epithelial cells lose polarity through cytoskeletal remodelling, individualize and gain motility. EMT is a critical process in embryonic development and wound healing, but it also plays a key role in fibrosis and cancer invasion. EMT and the reverse mesenchymal-epithelial transition (MET) are influenced not only by biochemical cues, but also by mechanical properties of the ECM \cite{Savagner2015, Poltavets2018}. 

Cells sense the mechanical properties of the ECM through their acto-myosin cytoskeleton at focal adhesions, which are based on integrin adhesion receptors. Integrins interact with the ECM through their extracellular domains and with the actin cytoskeleton through their cytoplasmic tails \cite{Ross2013}. The actin-integrin connection is mediated through talin and kindlin (reviewed in \cite{Calderwood2013}). Single integrins form small and transient junctions, but mechanical stimulation reinforces integrin adhesions by causing maturation into large focal adhesions. Upon mechanical stimulation, talin and kindlin underdo conformational changes that expose cryptic binding sites for additional cytoskeletal and signalling proteins \cite{Ross2013}. Mechanical stimulation further reinforces focal adhesions by inducing actin polymerization\cite{Puleo2019}. Variations in the biochemical composition and physical properties of the ECM can elicit different 3D cell migration modes characterized by different amounts of cell-ECM adhesion \cite{Petrie2016, Friedl2010}. Highly crosslinked and dense matrices elicit lobopodial migration, characterized by a high number of focal adhesions and high actomyosin contractility. Less dense, fibrous environments elicit mesenchymal migration with a characteristic front-to-rear gradient of focal adhesions. In low-confining areas that lack adhesion sites, cells depend on bleb formation to drive themselves forward, a mechanism that does not require focal adhesions \cite{Petrie2016}.

Not much is known about the role of cytoskeletal crosstalk in migration plasticity. Microtubules are likely involved through their feedback interactions with actin near focal adhesions. It was recently shown that higher substrate rigidity promotes microtubule acetylation through the recruitment of $\alpha$-tubulin acetyltransferase ($\alpha$ TAT) to focal adhesions by talin \cite{Seetharaman2022}. In turn, microtubule acetylation tunes the mechanosensitivity of focal adhesions by promoting the release of GEF-H1 from microtubules to activate RhoA and thereby promote actomyosin contractility \cite{Seetharaman2022}. In breast cancer cells, actin-microtubule crosstalk near focal adhesions via the scaffolding protein IQGAP1 was demonstrated to promote invasion in wound healing and transwell assays \cite{Mataraza2003}. In fibrosarcoma cells, the microtubule-destabilizing protein stathmin was shown to influence migration mode switching \cite{Belletti2003}. Increased stathmin activity, and as a result less stable microtubule networks, promoted amoeboid-like migration, while phosphorylation of stathmin led to a more elongated migratory phenotype. Besides crosstalk near focal adhesions, microtubules can also influence migration mode switching through mechanical effects. In confined or compressed cells, microtubules are stabilized through CLASP2 localization to the lattice, providing a mechanosensitive pathways for cells to adapt to highly constricting environments \cite{Li2023-kp}. 

The intermediate filament protein vimentin is considered a key cellular plasticity regulator and marker of tumor cell malignancy, especially based on its general upregulation in EMT and in motile cancer cells \cite{Ridge2022-ak}. Carcinoma cells in addition express integrin $\alpha$6$\beta$4, which recruits vimentin to focal adhesions through its binding to plectin, promoting a 3D invasive phenotype switch \cite{Qi2022}. Together with nuclear lamins, vimentin contributes to migration plasticity through regulation of nuclear deformation, for instance allowing for a transition towards amoeboid-like and faster migration in Hela cells and melanoma cells when encountering confinement \cite{Wang2023, Lavenus2020}. Moreover, cells migrating under high confinement use their nucleus as a piston to squeeze through small pores. This complicated pulling mechanism is regulated through crosstalk between the vimentin cage around the nucleus and actomyosin in front of the nucleus \cite{Petrie2014-zo}. 

\subsection{Cell-free reconstitution studies}

Understanding the molecular basis of migration plasticity is an enormous challenge since integrin-based matrix adhesions contain over 100 types of molecules that are potentially mechanosensitive \cite{Schiller2011-cv}. One of the first mechanotransduction events during adhesion maturation is stretching of talin, followed by vinculin binding and activation. This core process was elegantly reconstituted by overlaying a network of actin-myosin bundles mimicking stress fibers on a talin-micropatterned surface \cite{Ciobanasu2014}. It was shown that direct binding of the contractile actin-myosin network to talin was sufficient to stretch the protein and induce the association and activation of vinculin. Talin binding facilitates vinculin activation by allosterically weakening the head–tail interaction that keeps it in an auto-inhibited conformation \cite{Franz2023-fw}. Exposure of the actin-binding tail induces a positive feedback that reinforces the connection with actin \cite{Ciobanasu2014}. Using the same assay, it was shown that activated vinculin can interact with Arp2/3 complex-mediated branched actin networks and modify their organization by crosslinking actin filaments into bundles \cite{Boujemaa-Paterski2020}. This is likely an important step towards focal adhesion maturation. Single-molecule studies showed that vinculin forms a directionally asymmetric catch bond with F-actin \cite{Huang2017}. In this way vinculin can organize the polarity of the actin cytoskeleton and contribute to front-rear asymmetry in migrating cells. Recently the interaction of integrins, talin and kindlin, another major focal adhesion regulator, was reconstituted on giant unilamellar vesicles \cite{Pernier2023-tt}. It was shown that phosphoinositide-rich membranes recruit talin and kindlin, which then cause the formation of large integrin clusters that can recruit actin-myosin. Another study showed that membrane-bound talin can also activate vinculin and the two proteins together can link actin to the membrane \cite{Kelley2020-xt}.

Cell-free reconstitution studies suggest that the actin cytoskeleton itself also contains proteins that mediate mechanotransduction. An example is filamin A (FLNA), a large multi-domain scaffolding protein that cross-links actin filaments and binds numerous proteins via cryptic binding sites along its length. Using reconstituted actin networks crosslinked with FLNA, it was shown that mechanical strain on the FLNA crosslinks alters its binding affinity for its binding partners\cite{Ehrlicher2011}. Both externally imposed bulk shear and contraction by myosin-II increased binding of the cytoplasmic tail of $\beta$-integrin while it weakened binding of FilGAP, a GTPase that inactivates Rac. Mechanical strain on FLNA can thus stabilize extracellular matrix binding and at the same time influence actin dynamics through Rac activity.

\section{The road ahead}

\begin{figure}[t]
\centering
\includegraphics[scale=0.30]{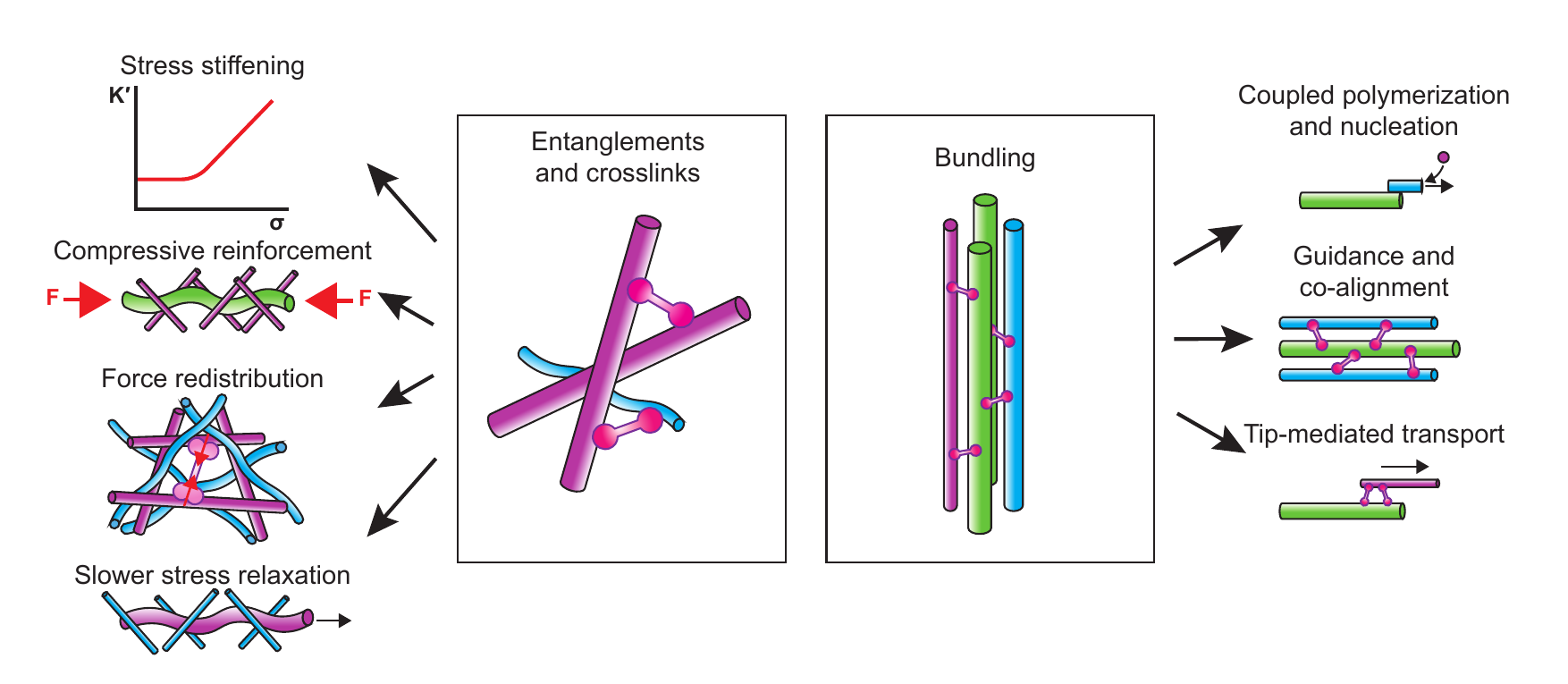}
%\captionsetup{margin=0.5cm}
\caption{Schematic of the two general cytoskeletal crosstalk mechanisms and their effect on cytoskeletal biophysics. $Entanglements$ and $crosslinks$ regulate stress stiffening, compressive reinforcement, force distribution and slower stress relaxation. $Bundling$ regulates coupled polymerization and nucleation, guidance and co-alignment, and tip-mediated transport of filaments. Actin (magenta), microtubules (green), intermediate filaments (cyan) and crosslinkers (pink).}
\label{mech}
\end{figure}

Cytoskeletal crosstalk is increasingly recognized as a major determinant of cell migration. Coupling between the actin, microtubule and intermediate filament cytoskeleton influences cell migration by regulating cell deformability, contractility, front-rear polarity and migration plasticity. Coupling of filaments through entanglements, crosslinking and bundling of filaments regulates a variety of mechanisms that mediate cellular mechanics and cytoskeletal dynamics involved in these migration strategies (Figure \ref{mech}). To complicate matters, there is growing evidence that a fourth cytoskeletal protein family, the septins, also strongly impacts cell migration. Septins are well-equipped to mediate cytoskeletal crosstalk since they can bind to the cell membrane, actin and microtubules \cite{Spiliotis2020-bf}. Recent research demonstrated roles of septins in mesenchymal and amoeboid single-cell migration \cite{Nakos2022-va, Zhovmer2024} and in the regulation of endothelial and epithelial cell-cell adhesion \cite{Kim2022, Wang2021}. Elucidating the biophysical mechanisms by which cytoskeletal crosstalk regulates cell migration is challenging due to the enormous molecular complexity of the cell and feedback between mechanical forces and biochemical signaling. Cell-free reconstitution provides a valuable complement to live-cell studies because it simplifies the challenge of separating biochemical and physical contributions to cytoskeletal crosstalk. However, there is still an enormous gap gap between the complexity of cells and the simplicity of reconstituted systems. How can this gap be bridged? 

\begin{figure}[h!]
\centering
\includegraphics[scale=0.30]{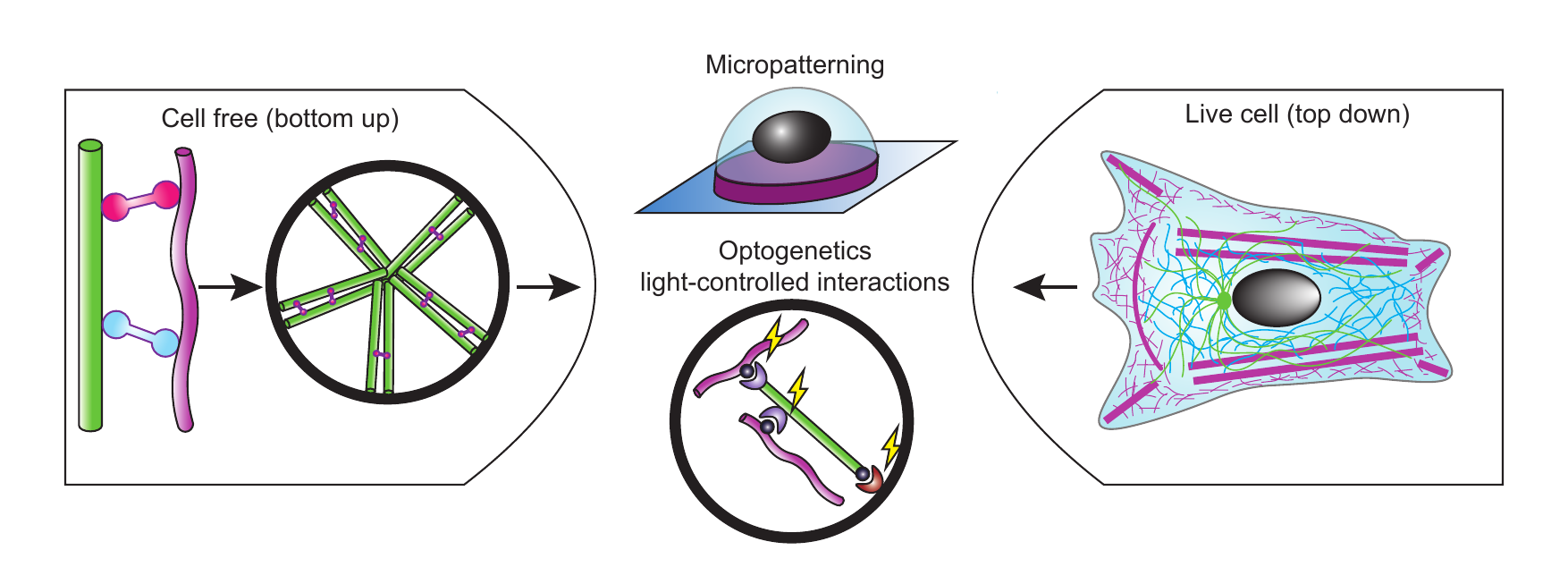}
%\captionsetup{margin=0.5cm}
\caption{The proposed route to bridge the gap between live cell (top down) and cell free (bottom up) approaches in the research of cytoskeletal crosstalk in cell migration. $Micropatterning$ and $optogenetic$ $tools$ can be used to manipulate cytoskeletal interactions and help to bridge the gap between the two research approaches. Actin (magenta), microtubules (green), intermediate filaments (cyan) and crosslinkers (pink).}
\label{outlook}
\end{figure}

We propose different routes to bridge this gap (Figure \ref{outlook}). One obvious direction to bridge this gap is to enhance the complexity of cell-free assays. Simple assays combining cytoskeletal filaments without any accessory proteins showed that steric interactions alone suffice to give rise to nonlinear stiffening and enhanced compressive strength. There is some evidence that intermediate filaments directly interact with actin filaments and microtubules via electrostatic interactions \cite{Esue2006, Schaedel2021}, but this could be an artefact of \textit{in vitro} conditions. Single-molecule measurements of filament interactions within cells or cell lysates could shed light on this issue. An important next step to bridge the gap to the cell is to incorporate accessory proteins that mediate cytoskeletal coupling. Several studies have shown that crosslinking via cytolinkers is sufficient to give rise to cytoskeletal filament co-alignment and mechanical synergy (e.g. \cite{Preciado_Lopez2014-af, Petitjean2023}). More detailed investigations of the effects of cytolinkers, both \textit{in vitro} and in cells, will be important to delineate their roles in cytoskeletal co-organization, mechanical synergy, and mechanotransduction. In the cell, cytoskeletal crosstalk is guided by geometrical constraints provided by the cell membrane. The membrane organizes the cytoskeleton through spatial confinement and by providing adhesion sites where cytoskeletal filaments are nucleated or anchored. Reconstitution experiments have begun to recapitulate these constraints by encapsulating cytoskeletal proteins inside cell-sized emulsion droplets or lipid vesicles, including actin/microtubule and actin/keratin composites \cite{Deek2018-hv, Vendel2020-sl}. These model systems could form a basis for reconstituting synthetic cells capable of migration. Adhesion-independent migration is probably easiest to reconstitute. Flow-driven confined migration was recently reconstituted, although based on a cell extract, so the minimal set of ingredients is not yet known \cite{Sakamoto2022-hi}. It will be interesting to incorporate microtubules and/or intermediate filaments in this assay to control cell polarity and mechanics. Mesenchymal migration is likely more challenging to reconstitute because it requires coordinated actin polymerization, contraction, and cell-matrix adhesion. Motility driven by actin polymerization has been successfully reconstituted on the outer surface of lipid vesicles (reviewed in \cite{Siton-Mendelson2016-ta}), but motility of vesicles with actin polymerization inside will require substrate adhesion. Surface micropatterning provides an interesting approach to impose polarized shapes to synthetic cells by forcing them to adapt to the pattern shape and size \cite{Fink2023-tr}. In addition, one can use light-induced dimerization to induce spatial patterning and symmetry breaking of cytoskeletal networks. Light-inducible dimers (LIDs) come from photoactivatable systems naturally occurring in plants and allow for reversible photoactivation \cite{Wittmann2020-se}. Recently it was for instance shown that microtubule-interacting proteins fused to optochemical dimerization domains can be used to drive symmetry breaking of microtubule networks inside emulsion droplets \cite{Bermudez2021-xb}. 

The opposite direction to bridge the gap between live-cell and cell-free studies is to tame the complexity of living cells. Some of the same techniques that can provide more control over cell-free systems can also provide control over the behavior of living cells. Surface micropatterning for instance allows one to confine cells to adhesive islands with precisely controlled geometries, forcing the cells to adopt prescribed shape and corresponding cytoskeletal organizations. Imaging many cells adhered on the same pattern greatly facilitates quantification of cytoskeletal crosstalk \cite{Goldman2014}. Moreover, micropatterning can be used to investigate how cytoskeletal interactions affect single-cell and collective cell migration dynamics \cite{Bruckner2021-te}. Light-inducible dimerization can be used to manipulate cytoskeletal interactions with high spatial and temporal control. It was for instance recently shown that F-actin can be crosslinked to microtubule plus ends by transfecting cells with an iLID-tagged EB-binding SxIP peptide and SspB-tagged actin-binding domains \cite{Adikes2018-cs}. This could be an interesting tool to systematically study the crosstalk of microtubules with actin and intermediate filaments that takes place near cell adhesions. Finally, molecular tension sensors provide a very interesting tool to selectively interrogate mechanical interactions between cytoskeletal networks. Tension sensors consist of two fluorescent proteins separated by a peptide with a calibrated mechanical compliance. Under strain, the fluorescent proteins are separated, decreasing fluorescence energy transfer (FRET) between them. By embedding a tension sensor in the actin crosslinker FLNA, it was recently shown that molecular tension can be measured within the actin cytoskeleton  \cite{Amiri2023-dr}. It will be interesting to use a similar approach to measure tension within the intermediate filament cytoskeleton and test force transmission between the actin and intermediate filament cytoskeleton.

%\bibliography{Review_bib2}

\begin{acknowledgments}
We gratefully acknowledge funding from the VICI project \textit{How cytoskeletal teamwork makes cells strong} (project number VI.C.182.004) and from the OCENW.GROOT.20t9.O22 project \textit{The Active Matter Physics of Collective
Metastasis}, both financed by the Dutch Research Council (NWO).
\end{acknowledgments}

\section{Data Availability Statement}
Not applicable.

\section{Bibliography}

%\nocite{*}
\bibliography{aipsamp}% Produces the bibliography via BibTeX.

\end{document}